\documentclass[aps,twocolumn,prb,reprint,amsmath,amssymb,showpacs,superscriptaddress]{revtex4-1}

\usepackage{graphicx}
\usepackage{dcolumn}
\usepackage{bm}
\usepackage{amsmath}
\usepackage{color}
\usepackage{mathrsfs}
\usepackage{float}
\usepackage{dsfont}
\usepackage{txfonts}
\usepackage{wasysym}
\usepackage{ifsym}

\makeatletter

\newcommand{\Rmnum}[1]{\expandafter\@slowromancap\romannumeral #1@}
\makeatother
\begin{document}

\title{Catalogue of Andreev spectra and  Josephson effects in structures with time-reversal-invariant 
topological  superconductor wires} 

\author{Liliana Arrachea}
\affiliation{International Center for Advanced Studies, Escuela de Ciencia y Tecnolog\'{\i}a, 
Universidad Nacional de San Mart\'{\i}n-UNSAM, Av 25 de Mayo y Francia, 1650 Buenos Aires, Argentina} 

\author{Alberto Camjayi}
\affiliation{Departamento de F\'{\i}sica, FCEyN, Universidad de Buenos Aires and IFIBA, 
Pabell\'on I, Ciudad Universitaria, 1428 CABA Argentina} 

\author{Armando A. Aligia}
\affiliation{Centro At\'omico Bariloche and Instituto Balseiro, CNEA, CONICET, 8400 S. C. de Bariloche, Argentina}

\author{Leonel Gru\~neiro}
\affiliation{International Center for Advanced Studies, Escuela de Ciencia y Tecnolog\'{\i}a, 
Universidad Nacional de San Mart\'{\i}n-UNSAM, Av 25 de Mayo y Francia, 1650 Buenos Aires, Argentina} 


\begin{abstract}
We study all the possible different two terminal configurations of Josephson junctions containing wires of 
time-reversal invariant topological superconductors (TRITOPS) and ordinary superconductors, 
including combinations with an interacting  quantum dot between both wires in the junction. 
We introduce simple effective Hamiltonians
which explain the different qualitative behaviors obtained. We analyze a wide range of phenomena, 
including occurrence and quenching of the so called  $0-\pi$ transition,
anomalous periodicity and jumps of the Josephson current as a function of the phase difference, and 
finite Josephson current in the absence of magnetic flux.
\end{abstract}

\date{\today}

\maketitle
\section{ Introduction}
Topological states of the matter is an ubiquitous topic of several areas of physics, including the communities 
of solid state physics,\cite{solid}, photonic  sciences,\cite{photo} and cold-atoms.\cite{coldat} 
The branch of topological superconductivity is paramount to the field of quantum computation, 
after the seminal idea by Kitaev\cite{kitaev} and coworkers\cite{kitaev-qc} of exploiting the non-abelian nature 
of anyons. Low-dimensional topological superconductors are promising candidates for realizing topological q-bits, 
since they host  Majorana zero-modes as edge states \cite{kitaev-model,ali}. This motivated a number of proposals to 
produce the topological phases in structures of quantum wires proximity-coupled to macroscopic superconductors as 
q-bit platforms. A prominent one is the model proposed in Refs.~\onlinecite{wires1,wires2}, which is under 
active experimental investigation.\cite{wires-exp1,wires-exp2,wires-exp3,wires-exp4}

The Kitaev model\cite{kitaev-model} and several other models for topological superconductivity,\cite{wires1,wires2}
are based on ingredients that break time-reversal symmetry. In contrast, there is another family, the TRITOPS,  
which belongs to the DIII class and hosts zero-energy edge excitations that appear in Kramers 
pairs. So far there are no experimental realizations of this topological phase, although it is receiving significant theoretical attention 
in the last years.\cite{qi,fu-berg,ryu,santos,wong,dumi1,zkm,ady,haim,yuval,klino,schra,tritops-bt,tritops-ort,chung,yaco,mat,tritops-ber,mellars,gong,nos,entangle,review,sch-fu,mash,par,tri-os,jorg}
An interesting characteristic, of these end modes is the fact that they have fractional spin projection.\cite{ady}
This property can be relevant for their detection.\cite{tritops-ber,nos,entangle,sch-fu}  Recently a universal gate set using TRITOPS has been 
proposed.\cite{schrade}

One of the most relevant and clear signatures of the superconducting phase is the Josephson effect, 
which takes place in superconductor structures with an annular shape threaded by a magnetic flux. 
The behavior of the
generated Josephson current as a function of the phase $\phi = 2\pi \Phi / \Phi_0$, 
where $\Phi$ is the flux and $\Phi_0= h/2e$ is the superconducting flux quantum, 
constitutes a valuable tool to unveil interesting 
physics related to the nature of the superconductor or to the junction itself. 
In the case of ordinary superconductors, the current-phase relation has a periodicity of $2 \pi$. 

 In the case of junctions of topological superconducting wires, the periodicity of the 
Andreev spectrum is $4 \pi$ ($2 \pi$) if the electron parity of the system is (is not) conserved and there is a level crossing at $\phi=\pi$, 
which leads to a peculiar behavior of the Josephson junction.\cite{kitaev-model,trs1} 
Josephson junctions containing one-dimensional topological superconductors with broken time-reversal (TR) symmetry 
have been studied in several works. Since the proposal by Fu and Kane  for   the realization of the topological phase in a quantum spin Hall system with a magnetic island, in proximity with a superconductor, the behavior of the Josephson current  was suggested as a way to detect the topological phase,\cite{trs2} and this system was 
later analyzed in other works.\cite{trs3,trs4} Dynamical effects in  Josephson junctions with topological 
wires with broken TR  were also analyzed,\cite{trs3,trs5,trs6} 
as well as other configurations including quantum dots with many-body interactions and multiple 
terminals.\cite{trs7,trs8,trs9,trs10,trs11,trs12,trs13,trs14,trs15}

The Josephson current in junctions of ordinary (non-topological) time-reversal invariant superconductors with an embedded interacting quantum dot  (QD) has the same periodicity 
with the magnetic flux as the direct junctions without QD. Interestingly,  they exhibit the so 
called $0-\pi$ transition.\cite{0-pi-1,0-pi-2,0-pi-3,0-pi-4,0-pi-5,0-pi-6,0-pi-7,0-pi-8,0-pi-9,q-dots}  This  implies the inversion of the sign
of the Josephson current, as the control parameters change and modify the occupancy and the net spin of 
the QD, which is  a consequence of the many-body interactions and is originated in the competition of the Kondo effect with the superconducting pairing. When the former one is dominant, the system is in the $0$ phase, while when the latter turns to dominate, the transition to the $\pi$-phase takes place.
 When the junctions are made of TRITOPS wires  with an embedded interacting QD, and preserve at least one 
component of the spin,  the peculiarity is
the quenching of the $0-\pi$ transition, as discussed in Ref. \onlinecite{nos}. This is due to the screening 
of the magnetic moment of the quantum dot by a combination of the zero-energy modes 
at both sides of the quantum dot. 
This generates a correlated ground state and the Josephson current displays a $0$-phase behavior, irrespectively of the strength of the many-body interactions in the QD.
Other remarkable feature is the existence of  discontinuities with large jumps in the Josephson current  at both $\phi=0,\pi$ in junctions with one TRITOPS and another ordinary superconducting wire. \cite{chung,mellars}

The aim of the present work is the analysis and comparison of the Andreev spectrum and Josephson current of all 
the possible configurations containing one or two TRITOPS wires with spin-orbit interaction. 
We consider direct junctions and junctions with an embedded interacting QD between both wires.
For the case of two TRITOPS wires,  we consider the effect of different directions of the spin-orbit interaction 
of the two wires. We focus only on configurations which are time-reversal invariant  at $\phi=0,\pi$. 

The work is organized as follows. Section II contains the description of the theoretical model for the wires 
and the effective Hamiltonian used to analyze the subgap Andreev spectrum. 
Results are presented in section III. Section IV is devoted to summary and conclusions.

\section{Model}
The full structure is modeled by the Hamiltonian $H=\sum_{\alpha} H_{\alpha} + H_J$, 
where $\alpha=L,R$ labels the two wires, and $H_J$ is the Hamiltonian for the junction. 
We describe the different pieces separately
\subsection{Wires}
Each wire of the structure is modeled by the following Hamiltonian 
\begin{eqnarray}\label{wire}
H_{\alpha} &= &\sum_{i,j } \left[ \psi^{\dagger}_{\alpha,i} \;  h^{\alpha}_{ij} \;\psi_{\alpha,j}   +  \psi^{\dagger}_{\alpha,i} \; {\Delta}^{\alpha}_{ij} \; \psi^{\dagger}_{\alpha,j} \right] + {\rm H. c.} , \nonumber \\
 h^{\alpha}_{ij} &= & 
- \left(t_{\alpha}  + i  \lambda_{\alpha}  {\bf n}_{\alpha} \cdot \boldsymbol{\sigma} \right) \delta_{j,i+1} 
-  \mu_{\alpha} 
\delta_{i,j}  \nonumber \\
 {\Delta}^{\alpha}_{ij}  & =&    \left(  \tilde{\Delta}_{\alpha} \; \delta_{j,i+1} + \Delta_{\alpha} \; \delta_{i,j}  \right)\; i \sigma_y  
\end{eqnarray}
where we have introduced the spinor $\psi^{\dagger}_{\alpha,j} = \left( c^{\dagger}_{\alpha,j,\uparrow}, c^{\dagger}_{\alpha,j,\uparrow}\right)$ and $\boldsymbol{\sigma} =\left(\sigma_x,\sigma_y,\sigma_z\right)$ are the three Pauli matrices. 
The terms of the normal part of the Hamiltonian are the
nearest-neighbor hopping
$ t_{\alpha}$,   and the spin-orbit interaction $\lambda_{\alpha}$ in  the direction of the vector ${\bf n}_{\alpha}$.
The pairing potential has a local
s-wave component $ \Delta_{\alpha} $, as well as an extended-s-wave 
component
$\tilde{\Delta}_{\alpha}$ and $\mu_{\alpha}$ is the chemical potential.

The Hamiltonian of Eq. (\ref{wire}) 
has time-reversal symmetry and corresponds to the model  introduced by Zhang-Kane-Mele in Ref. \onlinecite{zkm}
for TRITOPS (DIII class)  in one dimension. The key ingredients for the topological phase of class DIII are 
the spin-orbit interaction $\lambda_{\alpha}$ and the extended
s-wave pairing $\tilde{\Delta}_{\alpha}$. The local s-wave pairing $\Delta_{\alpha}$ induces ordinary 
superconductivity. As discussed in Ref. \onlinecite{zkm}, the TRITOPS phase takes place within the range of 
chemical potentials
satisfying $|\mu_{\alpha} - \epsilon_{0,\alpha} | < \epsilon_{m,\alpha}$, 
being $\epsilon_{0,\alpha}=\left(t_{\alpha} \Delta_{\alpha}\right)/\tilde{\Delta}_{\alpha}$ and 
$\epsilon_{m,\alpha}= 2 |\lambda_{\alpha}| \sqrt{1- \Delta_{\alpha}^2/\left(4 \tilde{\Delta}_{\alpha}^2 \right) }$.
This topological phase hosts Kramers pairs of end modes localized at the left and right  edge of the wire, 
represented by 
fermionic operators $\gamma_{{\alpha},s} ,\; \tilde{\gamma}_{{\alpha},s} $, with $s=\pm$, satisfying 
(see Ref. \onlinecite{entangle}) 
\begin{equation}\label{ztri}
\gamma_{{\alpha},+}^{\dagger} = i \; \mbox{sgn} \left( \lambda_{\alpha} \tilde{\Delta}_{\alpha} \right) \gamma_{{\alpha},-},\;\;\;\;\;\;\;
\tilde{\gamma}_{{\alpha},+}^{\dagger} =-i \; \mbox{sgn} \left(  \lambda_{\alpha} \tilde{\Delta}_{\alpha} \right) \tilde{\gamma}_{{\alpha},-}.
\end{equation}
Here, $+,\; -$ denotes parallel or antiparallel orientations of the spin along the direction ${\bf n}_{\alpha}$.

 
 \subsection{Junction}
 We consider two types of junctions. (i) A direct junction, described by a tunneling Hamiltonian between the two wires,
 \begin{equation}
 H_{\rm J, dir} = t_J \sum_{s} \left( e^{i \phi /2} c^{\dagger}_{L,s} c_{R,s} + \rm{H.c.} \right).
 \end{equation}
 (ii) A junction containing an interacting quantum dot embedded between the two wires
  \begin{eqnarray}
 H_{\rm J, dot} & = &  \sum_{s} \left(t_L e^{i \phi/4} c^{\dagger}_{L,s} d_s+ t_R e^{i \phi/4} d^{\dagger}_s c_{R,s} + \rm{H.c.} \right) + H_ d, \nonumber \\
 H_d & = & \varepsilon_{d} \left( n_{d,\uparrow} + n_{d,\downarrow} \right) + U n_{d \uparrow} n_{d \downarrow}.
 \end{eqnarray}
 Here, $\phi=2 \pi \Phi/\Phi_0$, where $\Phi$ is the total magnetic flux and the index ${\alpha}=L,R$ corresponds to the end site of the $L, R$ wire, which intervenes in the connection of the junction.

\subsection{Configurations and symmetries}\label{conf-sim}
The different configurations are the following: S-S, TRITOPS-TRITOPS, S-TRITOPS. 
The non-interacting case can be analyzed by means of exactly diagonalizing the Hamiltonian of the wires. 
In order to analyze the configurations with the interacting QD, we resort to low-energy effective
Hamiltonians. As we will see, such effective Hamiltonians are also useful to understand most of the relevant physical 
properties of direct junctions.

Each wire separately has charge conjugation, time-reversal, mirror symmetry 
(the spin projection in the direction of the spin-orbit coupling is conserved), chiral. 
When they are connected having different orientations of the spin-orbit interaction, 
the two latter symmetries are broken. For a flux different from zero or half a superconducting
flux quantum ($\phi \neq 0,\pi$), also time-reversal symmetry is broken.
For the whole system the fermion parity is a conserved quantity.

\section{Low-energy effective Hamiltonians}
\subsection{S-S}\label{sec:s-s}
Following Ref. \onlinecite{0-pi-1}, we define the effective low-energy Hamiltonian
\begin{equation}
 H^{\rm eff}_{S-S}  = \sum_{\alpha} \left( \Delta_{\alpha} c^{\dagger}_{\alpha, \uparrow} c^{\dagger}_{\alpha, \downarrow} +  {\rm H. c.}\right)+ H_J,
 \end{equation}
 with $H_{\rm J} \equiv  H_{\rm J, dir}$, for a direct junction or $H_{\rm J} \equiv  H_{J,\rm dot}$, 
for a junction with an embedded quantum dot. We discuss in Section IV the  validity of this effective Hamiltonian.

\subsection{TRITOPS-TRITOPS}\label{sec:tri-tri}
Following Ref. \onlinecite{nos}, we define an effective Hamiltonian where only the degrees of freedom 
associated to the zero modes of the topological wires are taken into account. 
For the case of a direct junction we have $H^{\rm eff}_{\rm TRITOPS-TRITOPS}= H^{\rm eff}_{\rm J,dir}$, 
and when a dot is between the wires $H^{\rm eff}_{\rm TRITOPS-D-TRITOPS}= H^{\rm eff}_{\rm J,dot}$.
Depending on the configuration we obtain
\begin{eqnarray}\label{heftt}
H^{\rm eff}_{\rm J,dir} & = & t_J \sum_{s=\uparrow,\downarrow} e^{i\phi/2} \tilde{\gamma}^{\dagger}_{L,s} \gamma_{R,s} + {\rm H. c.} ,\nonumber \\
H^{\rm eff}_{\rm J,dot} & = & t_J \sum_{s=\uparrow,\downarrow} \left( e^{i\phi/4} \tilde{\gamma}^{\dagger}_{L,s} d_s+ d^{\dagger}_s \gamma_{R,s} \right)+ {\rm H. c.} + H_d.
\end{eqnarray}
The operators $\gamma_{\alpha,s}$ and $ \tilde{\gamma}_{\alpha,s}$ are related to the Bogoliubov operators representing the zero modes defined in Eq. (\ref{ztri}) through
the transformation
\begin{eqnarray}
\left(\gamma_{\alpha,\uparrow}, \; \gamma_{\alpha,\downarrow} \right)^T& = &  U_{\alpha} \; \left(\gamma_{{\alpha},+}, \; \gamma_{{\alpha},-} \right)^T, \nonumber \\
\left(\tilde{\gamma}_{\alpha,\uparrow}, \; \tilde{\gamma}_{\alpha,\downarrow} \right)^T& = &  U_{\alpha} \; \left(\tilde{\gamma}_{{\alpha},+}, \; \tilde{\gamma}_{{\alpha},-} \right)^T.
\end{eqnarray}
The unitary matrix relating the operators is defined from the orientation of the vector ${\bf n}_{\alpha} = \left(\sin \theta_{\alpha} \cos \varphi_{\alpha}, \sin \theta_{\alpha} \sin \varphi_{\alpha}, \cos \theta_{\alpha} \right)$, as follows
\begin{equation}\label{u}
U_{\alpha} = \left( \begin{array}{cc} 
 \cos\frac{\theta_{\alpha}}{2} &  -\sin\frac{\theta_{\alpha}}{2} e^{i \varphi_{\alpha}} \\
 \sin\frac{\theta_{\alpha}}{2} e^{-i \varphi_{\alpha}} &   \cos\frac{\theta_{\alpha}}{2}.
 \end{array} \right).
\end{equation}
Without loss of generality we set ${\bf n}_R $ along the $z$-direction and $\mbox{sgn} \left( \tilde{\Delta}_{\alpha} \lambda_{\alpha} \right)>0$.
Denoting by $\tilde{\gamma}_{L,+}=\tilde{\gamma}$ and
$\gamma_{R,+}=\gamma$, we have
\begin{eqnarray}
\tilde{\gamma}_{L,\uparrow}&=& \cos \frac{\theta}{2} \tilde{\gamma} - i e^{i \varphi} \sin \frac{\theta}{2} \tilde{\gamma}^{\dagger},\;\;\;\;\;\;\;\;\;\;
\gamma_{R,\uparrow}= \gamma,\nonumber \\
\tilde{\gamma}_{L,\downarrow}&=&  e^{-i \varphi} \sin \frac{\theta}{2} \tilde{\gamma} + i \cos \frac{\theta}{2} \tilde{\gamma}^{\dagger},\;\;\;\;\;\;\;\;
\gamma^{\dagger}_{R,\downarrow}= i \gamma,
\end{eqnarray}
Substituting this transformation in Eqs. (\ref{heftt}), and using the relations 
of Eq. (\ref{ztri}), we get the following effective Hamiltonians
\begin{eqnarray}\label{tri-tri}
H^{\rm eff}_{\rm J, dir}  &=& t_0 \tilde{\gamma}^{\dagger} \gamma +
\delta_0 \tilde{\gamma} \gamma + {\rm H. c.}, \\
H^{\rm eff}_{\rm J, dot}  &=& H_L+
t_{\phi} d^{\dagger}_{\uparrow} \gamma 
 -i t_{\phi} d^{\dagger}_{\downarrow} \gamma^{\dagger}  + {\rm H.c.}+ H_d \nonumber \\
H_L & = & \sum_{s=\uparrow,\downarrow}\left(
t_{s} \tilde{\gamma}^{\dagger}d_{s} +  \delta_{s} \tilde{\gamma}d_{s} \right)
\end{eqnarray}
We have defined
\begin{eqnarray}\label{param}
 t_0 & =& 2 t_J \cos \frac{\theta}{2} \cos\frac{\phi}{2}, \;\;\;\;\;\;
\delta_0 = -  2 t_J \sin \frac{\theta}{2} \sin\frac{\phi}{2} \; e^{i \varphi},\nonumber \\
t_{\uparrow}& = & t_{\phi} \cos \frac{\theta}{2}, \;\;\;\;\;
t_{\downarrow}  =  t_{\phi} e^{ i \varphi} \sin \frac{\theta}{2}, 
\;\;\;\;\;
t_{\phi}  = t_J e^{i \phi/4},\nonumber \\
\delta_{\uparrow}  & = & i t_{\phi} e^{- i\varphi} \sin \frac{\theta}{2},\;\;\;\;\;\;\;\;\; \delta_{\downarrow}   =  - i t_{\phi} \cos \frac{\theta}{2}.
\end{eqnarray}
Notice that Eqs. (\ref{tri-tri}) are actually independent of the azimuthal angle $\varphi$
as it might be expected. In fact, the latter can be eliminated by the
following gauge transformation $\tilde{\gamma} \rightarrow e^{i\varphi/2}\tilde{\gamma}$, $\gamma \rightarrow e^{i\varphi/2}\gamma$,
$d_{\uparrow} \rightarrow d_{\uparrow} e^{- i\varphi/2}$, $d_{\downarrow} \rightarrow d_{\downarrow} e^{ i\varphi/2}$. 

If the transformation that relates the zero modes $\gamma_{\alpha,\sigma}$ with those entering Eq. (\ref{wire})
is known (for example numerically in 
large chains or analytically as in Ref. \onlinecite{entangle}), $t_J$ can be calculated explicitly. In the second 
Eq. (\ref{heftt}) we have assumed $t_L=t_R=t$ for simplicity. In general $t_J$ is smaller but of the order of $t$.
Since in the construction of the effective Hamiltonian the energies above the superconducting gap have been neglected,
the quantitative validity of the effective Hamiltonian 
is restricted to $t_J \ll |\tilde{\Delta}_{\alpha}|, |\Delta_{\alpha}|$. 
The neglected terms in the derivation of the effective Hamiltonian
are the hybridization of the zero modes of the opposite chain (in the case of direct junction) 
or of the degrees of freedom of the quantum dot (for the junction with QD),
with the high-energy quasiparticles above the gap in the direct junction. 
Such processes can affect the parameters of the effective Hamiltonian, but 
do not modify its form.

\subsection{TRITOPS-S}\label{sec:tri-s}
In this case, the effective Hamiltonian is a combination of the effective Hamiltonians previously formulated. Concretely,
$H^{\rm eff}_{\rm TRITOPS-S} = H_{\rm J} + H_S$. The Hamiltonian $H_J$  of the junction corresponds to  $H^{\rm eff}_{\rm J, dir}$ for a direct junction and
$H^{\rm eff}_{\rm J, dot} $ for a junction with a quantum dot. In the former case, it can be expressed as follows
\begin{equation}\label{tri-s-dir}
H^{\rm eff}_{\rm J, dir} =
t e^{i\phi/2} \left( \tilde{\gamma}^{\dagger}c_{R,+} -i  \tilde{\gamma}c_{R,-} \right)  + \Delta_R c_{R,+}^{\dagger} c_{R,-}^{\dagger} + {\rm H. c.},
\end{equation}
where we have defined $\left(c_{R,+},c_{R,-}\right)^T=U_L \left(c_{R,\uparrow},c_{R,\downarrow}\right)^T$, with $U_L$ given in Eq. (\ref{u}).  It is interesting to notice the explicit phase $\pi/2$
in the effective pairing (2nd term of the above Hamiltonian), which is a consequence of the relations of Eq. (\ref{ztri}) satisfied by the zero end modes. This effectively introduces a phase in the junction,
in addition to the one due to the magnetic flux.

Similarly, for the case of a quantum dot embedded in the junction, we can perform the transformation $\left(d_{+},d_{-}\right)^T=U_L \left(d_{\uparrow},d_{\downarrow}\right)^T$ and 
define $\left(c_{R,+},c_{R,-}\right)^T \equiv \left(c_{R,\uparrow},c_{R,\downarrow}\right)^T$. Then, the effective
Hamiltonian results
\begin{equation}\label{tri-s}
H^{\rm eff}_{\rm J, dot}  = H_d+
t_{\phi} \left( \tilde{\gamma}^{\dagger}d_{+} -i  \tilde{\gamma}d_{-} + \sum_{s=\pm} d^{\dagger}_{s} c_{R,s}  \right)
+ \; \Delta_{R} c_{R,+}^{\dagger} c_{R,-}^{\dagger} +
{\rm H.c.}
\end{equation}
where we are using
the same definition of $t_{\phi}$ as in Eq. (\ref{param}). 

Notice that in this configuration, the
effective Hamiltonians become independent of the orientation of the spin-orbit of the TRITOPS wire, as expected.

\section{Results}

We show results  calculated by the exact numerical diagonalization of the full Hamiltonian in wires of finite length $L$, along with the many-body spectrum calculated by the exact diagonalization of the effective Hamiltonians  introduced in Sections \ref{sec:s-s}, \ref{sec:tri-tri} and \ref{sec:tri-s}. The Josephson current is calculated from 
\begin{equation}\label{jeff}
J (\phi)= \frac{\partial E_0^{\rm eff}(\phi)}{\partial \phi},
\end{equation}
being $E_0^{\rm eff}(\phi)$ the ground-state energy of the many-body spectrum calculated with the effective Hamiltonian. In the case of wires with a finite length, it can be calculated from the Matsubara Green function as follows
\begin{eqnarray}\label{je}
J_{\rm dir} (\phi)&=&2 t_J \sum_{s} \lim_{\tau=0^-} \mbox{Im}\left[{\cal G}_{LR,s}(\tau)\right],\nonumber \\
J_{\rm dot} (\phi)&=&2 t_J \sum_{s} \lim_{\tau=0^-} \mbox{Im}\left[{\cal G}_{Ld,s}(\tau)\right],
\end{eqnarray}
being ${\cal G}_{LR,s}(\tau)= - \langle T_{\tau} \left[ c_{L,s} (\tau) c^{\dagger}_{R,s} (0) \right] \rangle$ and  ${\cal G}_{Ld,s}(\tau)= - \langle T_{\tau} \left[ c_{L,s} (\tau) d^{\dagger}_{s} (0) \right] \rangle$.
The latter can be evaluated from the results of the single-particle spectrum. We will focus on the limit of temperature $T=0$.

\subsection{Direct tunneling}
\subsubsection{S-S}
This junction has been analyzed in several works.  \cite{0-pi-1,0-pi-2,0-pi-3,0-pi-4,0-pi-5,0-pi-6,0-pi-7,0-pi-8,0-pi-9}
The main characteristic of the Andreev levels of this type of junction is
the existence of an energy gap, which leads to a smooth behavior of the Josephson current.  By comparing  the results  for the Josephson current 
obtained by the direct diagonalization of the Hamiltonian for the connected wires using Eq. (\ref{je}) with those obtained from the ground-state energy of the effective Hamiltonian using Eq. (\ref{jeff}), we verify the good qualitative agreement between 
the two approaches. 

\begin{figure}[h]
\begin{center}
\includegraphics[width=\columnwidth]{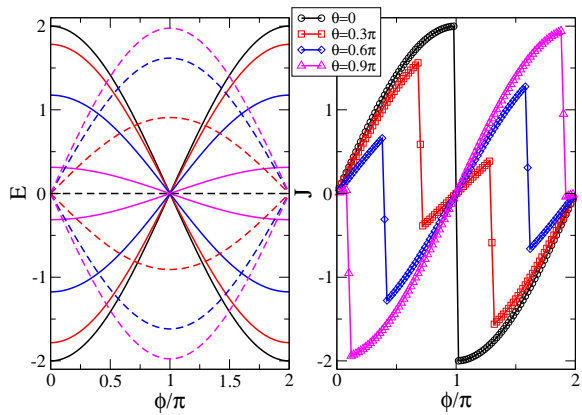}
\end{center}
\caption{Andreev spectrum  of a TRITOPS-TRITOPS junction with  direct tunneling between the wires calculated with the effective Hamiltonian (left) and the corresponding Josephson current calculated from
Eq. (\ref{jeff}) (right).
Solid and dashed lines correspond to states within the subspace with odd and even number of particles, respectively. Different plots correspond to different relative orientation $\theta$ in the direction of the spin-orbit coupling of the wires. 
Energies are expressed in units of $t$. }
\label{fig2}
\end{figure}

\subsubsection{TRITOPS-TRITOPS}\label{sec:tritops-tritops}
The solution of the effective Hamiltonian for the junction with direct tunneling between the wires is very simple. Details of the calculation of the many-body states are presented in Appendix \ref{appen}. 
The resulting spectra and Josephson current are shown in Fig. \ref{fig2}. We see that the effect of the relative orientation $\theta$ of the spin-orbit coupling of the two wires plays a crucial role in the behavior of the 
Andreev spectrum and the Josephson current. For wires with the spin-orbit coupling oriented along the same direction ($\theta=0$), the spectrum contains only two levels with flux-dependent energies,
which belong to the subspace of the effective Hamiltonian
with an odd number of particles, which cross at  $\phi=\pi$, while the states in the even subspace have zero energy, independently of the value of $\phi$. As a consequence of the level crossing, the Josephson current has a
discontinuity at $\phi=\pi$. This feature has been already discussed in Refs. \onlinecite{chung,review,mellars,nos} For wires with different orientation of the spin-orbit coupling 
($\theta \neq 0$), the states within the subspace with even number of particles of $H^{\rm eff}_{\rm TRITOPS-TRITOPS}$ become dispersive in $\phi$. Level crossing between states with different parities take place in the intervals
 $0 \leq \phi \leq \pi$ and $\pi \leq \phi \leq 2 \pi$, with the consequent discontinuities in the Josephson junction. 
 
In the limit where ${\bf n}_L$ and ${\bf n}_R$ form an angle $\theta=\pi$, the states within the subspace with odd number of particles do not disperse and have zero energy, while the ones in the even subspace fully determine the behavior of the Josephson current. 
An interesting feature is that, in this limit, the two  dispersive
Andreev states cross at $\phi=0$, leading
to a discontinuity with an abrupt jump in the Josephson current at zero flux. Such a behavior is quite peculiar and has been also pointed out in other Josephson junctions of topological superconductors with broken time-reversal 
symmetry. \cite{egger} Here, the origin is the effective phase in the junction, introduced by the different orientations of the spin-orbit interaction of the two wires. We note that when both directions coincide, the spin projection in this direction is a conserved quantity. However, this symmetry is lost in the more general case. 
From the exact solution of the effective Hamiltonian [Eqs. (\ref{ee}) and (\ref{eo})], 
we see that the role 
of $\theta$ and $\phi$ can be exchanged. 
In particular, changing $\theta$ from 0 to $\pi$ is equivalent to a shift in $\phi$ by $\pi$.  
This can be understood from the relations satisfied by the zero-modes given by the generalization of Eqs. (\ref{ztri}),
when the effect of the flux is shifted to a superconducting wire by a gauge transformation
[Eqs. (18), (21) and (26) of Ref. \onlinecite{entangle}].

\begin{figure}[h]
\vspace{1.cm}
\begin{center}
\includegraphics[width=\columnwidth]{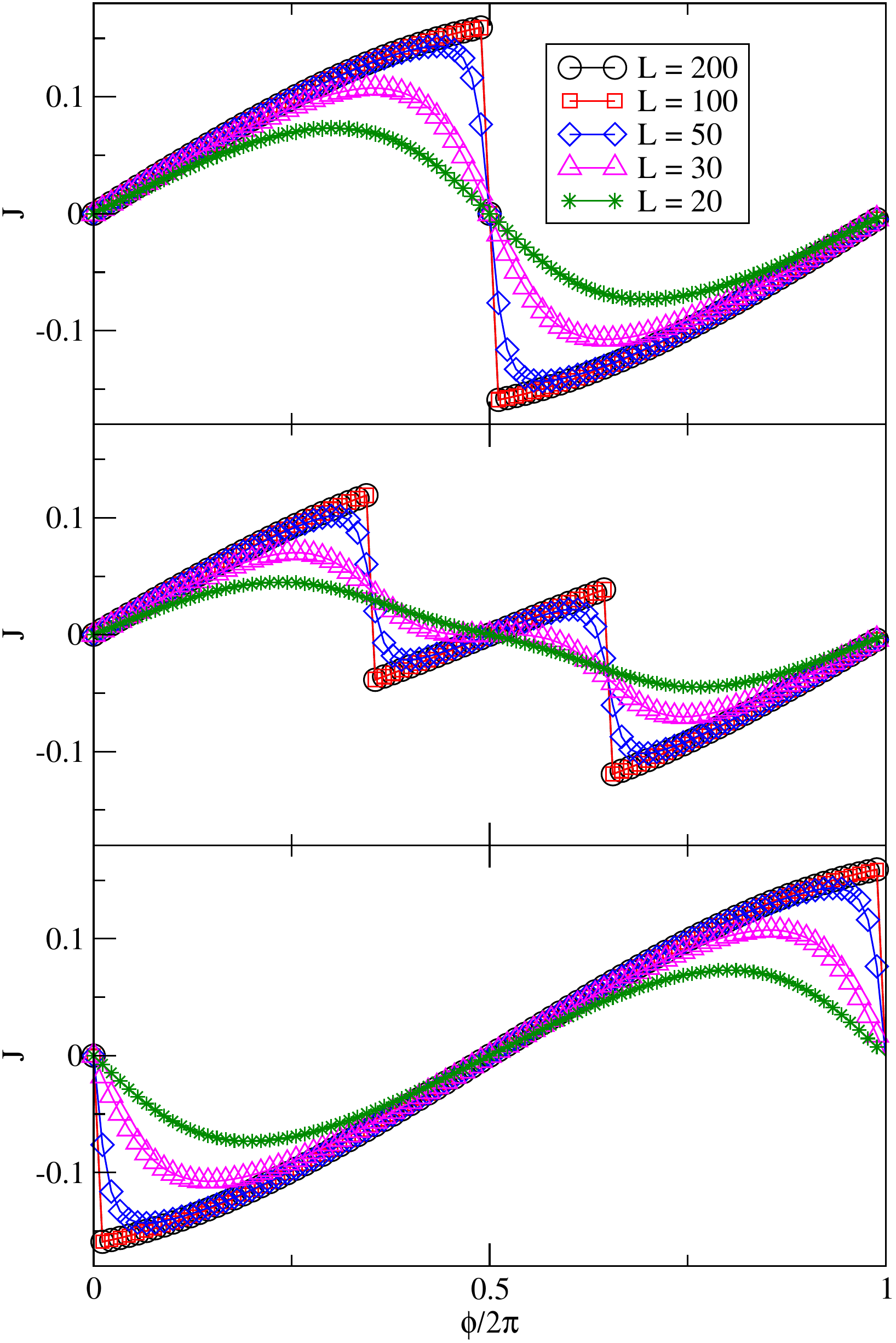}
\end{center}
\caption{Josephson current of a TRITOPS-TRITOPS junction with  direct tunneling between the wires for various orientations of the spin orbit of the left wire. Results correspond to exact diagonalization of finite chains.
Top, middle and bottom panels correspond to $\theta=0,0.3\pi,\pi$, respectively. Parameters are $\Delta=0,\; \tilde{\Delta} =0.2, \; \lambda=0.5,\; \mu=0$. Energies are indicated in units of $t$.}
\label{fig3}
\end{figure}
In Fig. 2, we show results for the Josephson current calculated on the basis of the exact diagonalization of 
wires with finite length $L$. We see that for the longer wires considered in the calculation, the results of the
Josephson current reproduce all the qualitative features already shown in Fig. 1, corresponding to the effective 
Hamiltonian. For short lengths, the end modes within each of the TRITOPS wires hybridize and the 
$J(\phi)$ response departs from the description of the effective Hamiltonian, 
where only the zero modes directly connected to the junction are taken into account.

\begin{figure}[h]
\begin{center}
\includegraphics[width=\columnwidth]{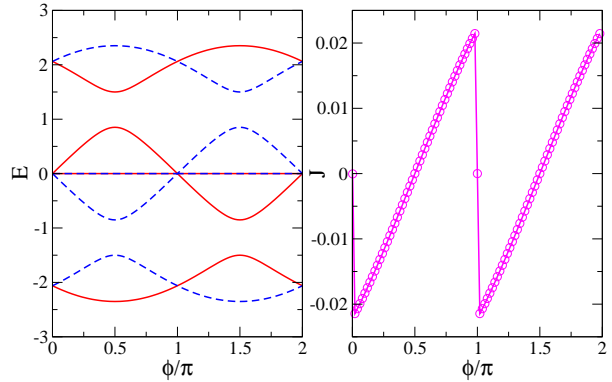}
\end{center}
\caption{Same as Fig. \ref{fig2} for a TRITOPS-S junction with  direct tunneling between the wires. We have considered $\Delta_R=1.5t$.}
\label{fig4}
\end{figure}

\subsubsection{TRITOPS-S}\label{rsec:tritops-s}
The solution of the effective Hamiltonian for the junction with direct tunneling between the wires is very simple. Details of the calculation of the many-body states are presented in Appendix \ref{appen}.
The resulting spectra and Josephson current are shown in Figs. \ref{fig4}. As mentioned in the derivation of the effective Hamiltonian, this configuration is independent of the orientation of the spin-orbit coupling.
The features to highlight are: (i) the discontinuity with a jump of the Josephson current at zero flux. This  is a consequence of the phase $\pi/2$ in the effective pairing along the junction, which can be traced back to the relation between the Bogoliubov
operators representing the zero modes, expressed in Eq. (\ref{ztri}). (ii) The other interesting feature is the level 
crossing at $\phi=\pi$. As it is clear from the corresponding effective Hamiltonian 
(see appendix A), both crossings at $\phi=0$ and $\phi=\pi$ are protected by fermion parity, since the two levels 
that cross belong to subspaces with different parity. 
(iii) It is also very interesting the fact that
the Josephson current seems to be approximately  periodic in $\pi$ instead of in $2 \pi$.
In fact, the low-energy effective Hamiltonian is exactly periodic in $\pi$ (see appendix A). In the case of wires of finite length, the coupling of zero modes modifies this picture, but the Fourier analysis of the 
Josephson current anyway presents a strong component associated to this half periodicity.

\begin{figure}[h]
\vspace{1.cm}
\begin{center}
\includegraphics[width=\columnwidth]{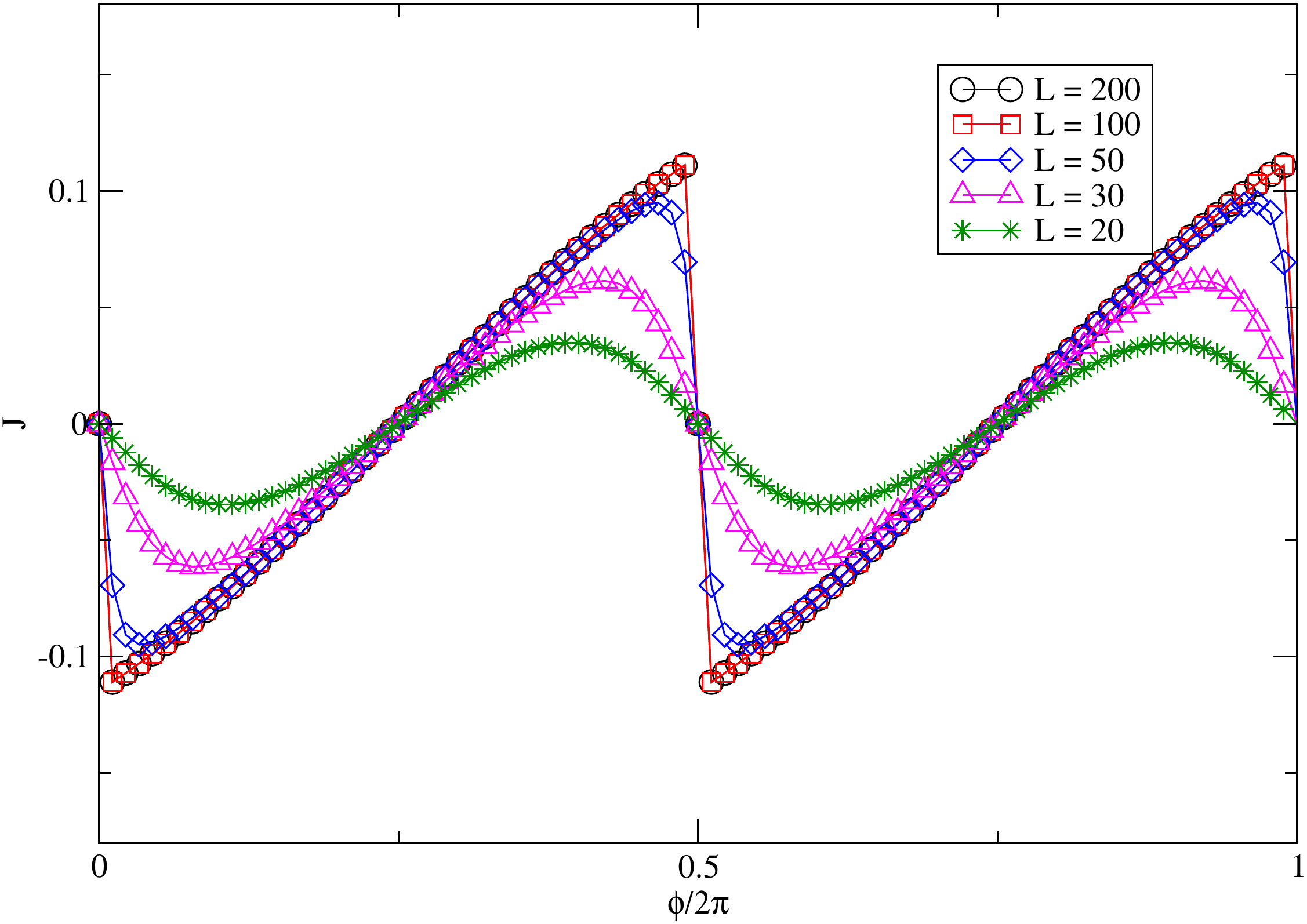}
\end{center}
\caption{Josephson current of a TRITOPS-S junction with  direct tunneling between the wires. 
Results correspond to exact diagonalization of finite chains. 
In the S wire $\Delta=0.5$.
Parameters for the TRITOPS wire are as in Fig. 3. }
\label{fig5}
\end{figure}

Results obtained by exact diagonalization of the wires are shown in Fig. \ref{fig5}, where the effect of the length of the wires can be appreciated.

\subsection{Junctions with embedded quantum dot}
\subsubsection{S-QD-S}

\begin{figure}[h]
\vspace{1.cm}
\begin{center}
\includegraphics[width=\columnwidth]{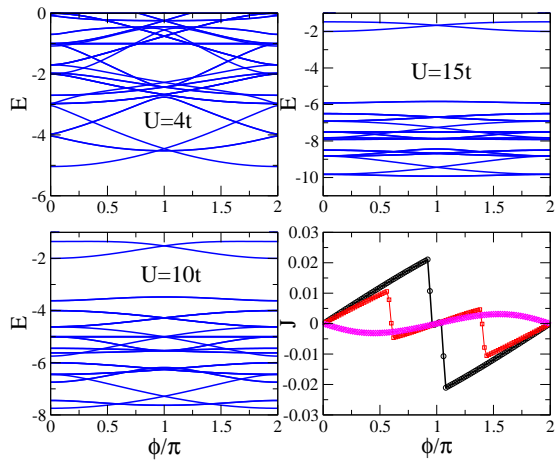}
\end{center}
\caption{Low-energy sector of the Andreev spectrum and Josephson current (bottom right) of an S-dot-S junction 
for different values of $U$, $\varepsilon_d=-U/2$ and $\Delta=t$. }
\label{fig6}
\end{figure}

The description of the leads --in which only one site is considered-- in the effective Hamiltonian
corresponds to the atomic limit,\cite{0-pi-1,0-pi-8} in which the superconducting gap $|\Delta_{\alpha}|$ is assumed to be
much larger then the hopping term $|t_{\alpha}|$. In spite of its simplicity, as we briefly discuss below
this approximation is able to describe qualitatively the $0-\pi$ transition .The occurrence of the $0-\pi$ transition is, precisely,
main characteristic of this junction. It takes place  as the parameters $\varepsilon_d$ and $U$ change with the quantum dot singly occupied.
This transition has been widely  discussed in the literature. \cite{0-pi-1,0-pi-2,0-pi-3,0-pi-4,0-pi-5,0-pi-6,0-pi-7,0-pi-8,0-pi-9}. In Fig. \ref{fig6} we review the evolution of the spectrum and the Josephson current as it takes place.

The $0-\pi$ transition is related to the  Kondo effect, which consists in the formation of a singlet between the localized spin at the QD and the spins of the electrons in the wires. This effect is characterized 
by the energy scale  set by the Kondo temperature $T_K$, 
which in the Kondo limit $t_\alpha \ll -\varepsilon_d$, $t_\alpha \ll \varepsilon_d+U$ and for $\Delta \rightarrow 0$ 
is $k_B T_K   \propto \exp [-1/(\rho J)]$ where  $\rho$ is the density 
of conduction states and 
$J=2  (t_L^2+t_R^2) U/[\varepsilon_d(\varepsilon_d+U)])$..
Hence, in the symmetric case $\varepsilon_d=-U/2$ as $U$ increases, the Kondo 
energy scale decreases. When  $k_B T_K \ll \Delta_R$, it becomes energetically non-convenient to
build the Kondo singlet between the localized electron and conductions electrons or holes at energy $\Delta_R$. Then,
the ground state becomes one with odd parity and an unscreened localized electron at the QD. The behavior of the ground-state energy as a function of $\phi$, 
changes from having a minimum at $\phi=0$ and a maximum at $\phi=\pi$ to the opposite situation, with the consequent 
change of sign in the Josephson current. The latter is, precisely,  known as the $0-\pi$ transition.

\begin{figure}[h]
\begin{center}
\includegraphics[width=\columnwidth]{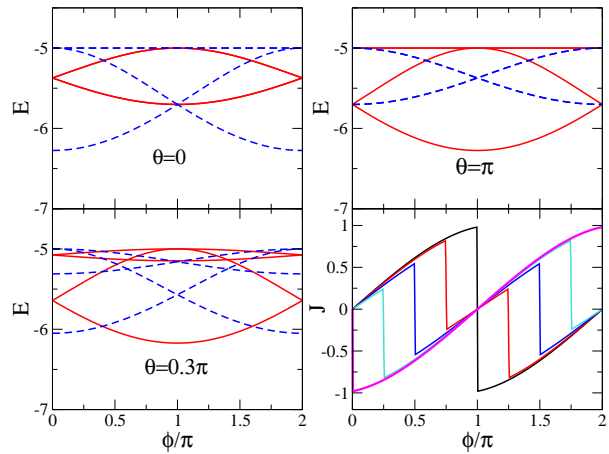}
\end{center}
\caption{Low-energy sector of the Andreev spectra of  TRITOPS-QD-TRITOPS junction. 
The Coulomb interaction and local potential of the quantum dot are, respectively,
$U=10t$ and $\varepsilon_d=-5t$. Different panels correspond to  different relative orientations 
of the spin-orbit couplings. Top right panel: the Josephson current for different values of 
$\theta=0$, $\pi/4$, $\pi/2$, $3 \pi/4$ and $\pi$.
Other details are the same as in Fig. \ref{fig2}.}
\label{fig7}
\end{figure}

\subsubsection{TRITOPS-QD-TRITOPS}
Results for the Andreev spectra of a junction with an embedded quantum dot calculated by diagonalizing the effective Hamiltonian are shown in Figs. \ref{fig7}. 
Only the low-energy sector of the spectrum is shown.
The case with the spin orbit of the wires oriented along the same direction ($\theta=0$) was previously 
analyzed in Ref. \onlinecite{nos} by means of a quantum Monte Carlo simulation to treat the interacting 
quantum dot. In addition an effective low-energy model of the Kondo type 
(constructed by a Schrieffer-Wolff transformation of the effective Hamiltonian) 
offered a simpler explanation of the main physics. 
The results shown in the top left panel of Fig. \ref{fig7} coincide with the latter picture. The spectrum is characterized by a four-fold degenerate
crossing at $\phi=\pi$. This is because, in addition to the crossing of two states belonging to the subspace with even number of particles, there is an additional crossing with a two-fold degenerate state belonging to the subspace with
odd number of particles. Consequently, the Josephson current presents a discontinuity at $\phi=\pi$. This feature has been already discussed in Ref. \onlinecite{nos}.

This behavior is strongly modified when different orientations of the spin-orbit coupling of the wires
are considered. As in the case of the direct junction analyzed in Fig. \ref{fig2}, crossings between states of the different subspaces 
take place within the intervals $0< \phi < \pi$ and $\pi< \phi < 2 \pi$. 
The jumps in the Josephson current take place at the crossing points. 
Also, as in the case of the direct junction analyzed in Section \ref{sec:tritops-tritops} in the limit of $\theta=\pi$, the crossing takes place at $\phi=0$, 
as a consequence of the phase between the zero end-modes given in Eq. (\ref{ztri}).
This can be explicitly seen in Fig. \ref{fig7}. 

Overall, the behavior of the Josephson current is practically unaffected by the many-body interaction at the quantum dot. As in the case analyzed in
Ref. \onlinecite{nos} there is no signature of the $0-\pi$ transition as a function of  $U$ and the occupancy of the QD.  This is because the low-energy spectrum is dominated by the zero modes, which hybridize to the QD 
to form a combined state akin to the Kondo singlet, irrespectively of the value of $U$. In the present case, 
the low-energy effective Hamiltonian provides the right qualitative description of the physics 
(high energy perturbative processes only introduce minor corrections to the parameters).

\begin{figure}[h]
\vspace{1.cm}
\begin{center}
\includegraphics[width=\columnwidth]{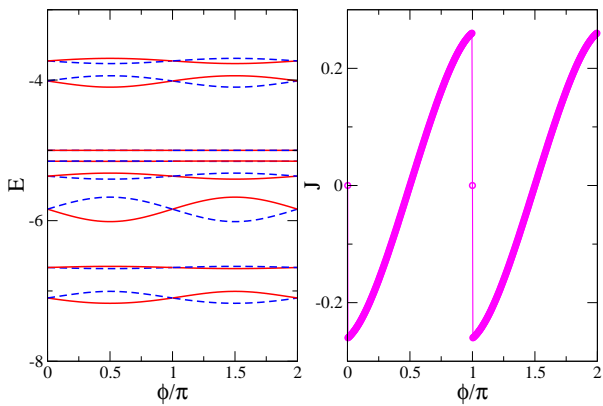}
\end{center}
\caption{Andreev spectrum (left) and Josephson current (right) of a TRITOPS-QD-S junction 
for $\Delta_R=1.5t$. Other  parameters  are the same as in Fig. \ref{fig7}.}
\label{fig8}
\end{figure}
\subsubsection{TRITOPS-QD-S}
The solution of the effective Hamiltonian of Eq. (\ref{tri-s}) for the junction between  TRITOPS and S wires  with the embedded quantum dot leads to the spectrum and Josephson current 
shown in Figs. \ref{fig8}. We see similar features as in the case of the direct coupling. 
In particular the features (i), (ii) and (iii) described in Section \ref{rsec:tritops-s} are also observed in the case
of the junction with the embedded QD. 

It is also interesting to highlight the absence of  $0-\pi$ transition in this configuration. 
Due to the crossing of states of different fermion-parity at flux $\pi$, there is always a 
relative maximum in the ground state energy $E(\phi)$ there. Therefore, the $0-\pi$ transition associated with 
a change from a maximum of $E(\phi)$ at $\phi=\pi$ to a minimum, accompanied by a change in fermion parity 
from even to odd 
as $U$ increases,  does not take place. 

In the limit  of  $\Delta_R \rightarrow \infty$, we can derive the effective low-energy Hamiltonian by recourse to a
Schrieffer-Wolff transformation, which reads
\begin{equation}
H_{\rm low} = J \left[\left(n_{d \downarrow} -n_{d \uparrow} \right) \tilde{\gamma}^{\dagger} \tilde{\gamma} + n_{d \uparrow} \right],
\end{equation}
with $J= - |t_\phi|^2 U /\left[\varepsilon_d \left(U+\varepsilon_d\right) \right]$. Therefore, as in the case of the QD connected to two TRITOPS wires, the low-energy states correspond to hybridizations of the zero modes with the QD, irrespectively
of the value of $U$. The ground state does not experience any qualitative change, which explains the lack of $0-\pi$ transition as $U$ changes.

 \section{Summary and conclusions}
 We have analyzed the Andreev spectrum and the Josephson current in different two-terminal configurations 
 containing one or two TRITOPS wires  with spin-orbit interaction.
 We have analyzed the possibility of different orientations of the spin-orbit coupling and the effect of 
 many-body interactions when a quantum dot is embedded into the junction between both wires.
 
 For TRITOPS-TRITOPS configurations we find that the phase introduced by the relative orientation of the 
 spin-orbit coupling of the wires plays a role, which is similar to the one due to the
 magnetic flux. In this way, level crossings and the consequent jumps of the Josephson current shift from 
 $\phi=\pi$ for parallel orientations of the spin orbit to $\phi=0$ to antiparallel ones.
 
 For TRITOPS-S junctions, we find an abrupt discontinuity with a jump of the  Josephson current at $\phi=0$ as in previous 
 works, as well as a strong component with periodicity of half the superconducting 
 flux quantum.\cite{chung,review,mellars} These features are, however, modified in wires of finite length due to the hybridization of the zero modes.
 
 For both cases, we find a quench of the $0-\pi$ transition in the presence of an interacting quantum dot in 
 the junction. The reason is the hybridization of the states localized at the quantum dot with
 the zero modes of the TRITOPS wires, that leads to a formation of a low-energy singlet. 
 We show that the zero mode of a single wire is enough to screen the localized state inhibiting the 
 transition to the $\pi$ phase. In contrast, for ordinary superconductors the formation of a Kondo singlet requires 
 taking quasiparticles with energy above the band gap.
 
 All these features can be explained in terms of simple low-energy effective Hamiltonians.
 
\section*{Acknowledgments}
LA thanks  Arbel Haim and Yuval Oreg for interesting discussions. We acknowledge support from CONICET, and UBACyT, Argentina  and  the Alexander von Humboldt
Foundation, Germany (LA). We are sponsored by PIP 112-201501-00506 (AA), PIP-RD 20141216-4905 
of CONICET (LA, AC and LG)  and PICT-2014-2049 (LA, AC and LG).

 \appendix
 
 \section{Solution of $H^{\rm eff}$ for a junction with direct tunneling}\label{appen}
\subsubsection{TRITOPS-TRITOPS}
 The many-body states of  $H^{\rm eff}_{\rm J, dir}$ given in Eq. (\ref{tri-tri}) can be easily constructed with the fermionic operators $\gamma$ and $\tilde{\gamma}$. These states are
 \begin{equation}
 |1\rangle = |0,0\rangle, \;|2\rangle = |1,0\rangle, \;|3\rangle = |0,1\rangle,\;|4\rangle = |1,1 \rangle,
 \end{equation}
 where the left and right entry denotes the occupation states of  $\gamma^{\dagger} \gamma$ and $\tilde{\gamma}^{\dagger}\tilde{\gamma}$, respectively.  $H^{\rm eff}_{\rm J, dir}$ conserves the parity. In the
 even subspace, the two eigenenergies are 
 \begin{equation}\label{ee}
 E^e_{\pm}(\phi)=\pm |\delta_0 (\phi)|= \pm 2 t_J \cos(\theta/2) \cos(\phi/2),
 \end{equation}
  while in the odd subspace they are 
  \begin{equation}\label{eo}
  E^o_{\pm}(\phi)=\pm t_0 (\phi)= \pm2 t_J \sin(\theta/2) \sin(\phi/2).
  \end{equation}
 
\subsubsection{TRITOPS-S}

We proceed in a similar way as in the previous section to construct the
many-body states of $H_{\mathrm{J,dir}}^{\mathrm{eff}}$ given in Eq. 
(\ref{tri-s-dir}). In the present case, the states have 3 entries, corresponding
to the number states of $\gamma ^{\dagger }\gamma $, $c_{R,\uparrow
}^{\dagger }c_{R,\uparrow }$, and $c_{R,\downarrow }^{\dagger
}c_{R,\downarrow }$. 
\begin{eqnarray}
&&|1\rangle =|0,0,0\rangle ,\;|2\rangle =|1,1,0\rangle ,\;|3\rangle
=|0,1,1,\rangle ,\;|4\rangle =|1,0,1\rangle ,  \notag \\
&&|5\rangle =|1,0,0\rangle ,\;|6\rangle =|0,1,0\rangle ,\;|7\rangle
=|0,0,1,\rangle ,\;|8\rangle =|1,1,1\rangle ,  \notag
\end{eqnarray}%
where the upper and lower lines correspond to the states of the even and odd
subspaces, respectively. Concerning the ordering of operators $|8\rangle
=\gamma ^{\dagger }c_{R,\uparrow }^{\dagger }c_{R,\downarrow }^{\dagger
}|0,0,0\rangle $, and the same ordering is used for the other states.

The resulting Hamiltonian matrix for the subspace with even number of
particles is 
\begin{equation}
H_{e}=\left( 
\begin{array}{cccc}
0 & 0 & \Delta _{R} & ite^{i\phi /2} \\ 
0 & 0 & 0 & 0 \\ 
\Delta _{R} & 0 & 0 & te^{-i\phi /2} \\ 
-ite^{-i\phi /2} & 0 & te^{i\phi /2} & 0 \\ 
\end{array}%
\right) ,  \label{he}
\end{equation}%
while for odd number of particles 
\begin{equation}
H_{o}=\left( 
\begin{array}{cccc}
0 & te^{i\phi /2} & 0 & \Delta _{R} \\ 
te^{-i\phi /2} & 0 & 0 & -ite^{i\phi /2} \\ 
0 & 0 & 0 & 0 \\ 
\Delta _{R} & ite^{-i\phi /2} & 0 & 0%
\end{array}%
\right) ,  \label{ho}
\end{equation}%
Note that the states $|2\rangle $ and $|7\rangle $ are eigenstates with
energy 0.

For even number of particles, the characteristic polynomial $P_{e}(E)=\det
(H_{e}-E)$ excluding the eigenstate $|2\rangle $ takes the form
\begin{equation}\label{pe}
P_{e}(E)=-E^{3}+(\Delta _{R}^{2}+2t^{2})E-2t^{2}\Delta _{R}\sin (\phi ).
\end{equation}

For odd number of particles, the characteristic polynomial $P_{o}(E)=\det
(H_{o}-E)$ excluding the eigenstate $|7\rangle $ is

\begin{equation}
P_{o}(E)=-E^{3}+(\Delta _{R}^{2}+2t^{2})E+2t^{2}\Delta _{R}\sin (\phi ).
\label{po}
\end{equation}

Comparing Eqs. (\ref{pe}) and (\ref{po}) one realizes that the spectrum for
odd number of particles coincides with that for even number of particles
with flux shifted by half a superconducting flux quantum ($\phi \rightarrow
\phi +\pi $). Furthermore, there is a crossing of the ground-state energies 
for even and odd number of particles at $\phi =0$ and $\phi =\pi $, at
energy with $E=-\sqrt{\Delta _{R}^{2}+2t^{2}}$, with the corresponding jumps
in the Josephson current.




\end{document}